\documentstyle[12pt]{article}
\newcommand{\beq}{\begin{equation}}
\newcommand{\eeq}{\end{equation}}
\newcommand{\beqa}{\begin{eqnarray}}
\newcommand{\eeqa}{\end{eqnarray}}
\newcommand{\beqar}{\begin{eqnarray*}}
\newcommand{\eeqar}{\end{eqnarray*}}
\renewcommand{\a}{\alpha}

\renewcommand{\O}{\Omega}

\newcommand{\norm}[1]{\raise.3ex\hbox{:}#1\raise.3ex\hbox{:}}

\newcommand{\Tr}{{\rm Tr}}

\newcommand\e{{\rm e}}

\def\A{{\cal A}}

\def\H{{\cal H}}
\def\Z{{\cal Z}}
\def\N{{\cal N}}
\def\W{{\cal W}}
\def\D{{\cal D}}
\def\O{{\cal O}}
\def\S{{\cal S}}
\def\T{{\cal T}}

\begin{document}
\begin{titlepage}
\rightline{\small IHES/P/98/70}
\vskip 5em

\begin{center}
{\bf \huge
Type IIB Matrix Theory at Two Loops}
\vskip 3em

{\large N. Hambli\footnote{email:
hambli@ihes.fr; hamblin@ictp.trieste.it\hfil}}
\vskip 1em

{\em  Institut des Hautes Etudes Scientifiques\\
      35, route de Chartres, 91440 - Bures-sur-Yvette, France}	

\vskip 1em

{\em ICTP, P. O. Box 586, 34014 Trieste, Italy} 

\vskip 4em

\begin{abstract}
The IKKT matrix model was proposed to be a non-perturbative
formulation of type IIB superstring theory. One of its
important consistency criteria is that the leading one-loop
$1/r^8$ effective interaction between a cluster of
type IIB D-objects should not receive any corrections
from higher loop effects for it to
describe accurately the type IIB supergravity results.
In analogy with the BFSS matrix model {\it versus} the
eleven-dimensional supergravity example, we show in this
work that the one-loop effective potential in the IKKT  
matrix model is also not renormalized at the two-loop
order. 

\end{abstract}
\end{center}

\end{titlepage}

\setcounter{footnote}{0}
\section{Introduction}

One of the remarkable consequences of the open string theory
description of D-branes [1] is the existence of a close
correspondence between the supergravity and supersymmetric Yang-Mills
theory (SYM) results for certain interactions of D-branes. This
surprising idea emerges in its simplest form from the examination
of the low-energy dynamics of parallel Dp-branes [2]
which is found to be described by a $U(N)$ maximally supersymmetric 
Yang-Mills theory on the
$p+1$-dimensional worldvolume of the Dp-barnes. To understand
better this correspondence the authors of [3] described 
the interactions that arise between two D-branes in two 
ways. Either by considering the sum of all closed string
exchanges, or using the modular properties of string theory,
as the sum of one-loop amplitudes in the open strings ending 
on the branes in the siprit of Bachas's D-brane dynamics
calculation [4,5]. In general this constitues a 
relation between two different description within string 
theory, and requires keeping all the string modes for 
its validity. 
\smallskip 

As articulated first in [3], for {\it large} D-brane 
separations $r >> l_s$, where $l_s = \sqrt{\alpha'}$ is the 
string scale, the velocity-dependent interaction between the 
D-branes is most easily described by a supergarvity theory 
in terms of the {\it massless} closed string exchange. 
Whereas for substringy separations $r << l_s$, the effective 
interaction is best described by the dynamics of the 
{\it lightest} open strings stretching between the D-branes 
which is encoded in the SYM on the brane 
worldvolume\footnote{In the supergravity picture, the massive 
closed string modes induces exponentially falling additional
interactions. Whereas in the SYM picture, the massive open string
states contribute create higher derivative interactions on the
brane worldvolume field theory.}. This is not surprising since
the truncation of the sum over the string states in each description 
to the lightest modes of each type must be valid in very different 
regimes. In some special situations, however, 
with some residual supersymmetry (left unbroken by the velocity),
an approximate cancellation between the bosons and the fermions
persists [6] allowing for the decoupling 
of the massive string states, and hence a correspondence between 
the supergravity results based on the masless closed string exchange 
and the SYM theory on the brane worldvolume based on 
the lightest open strings stertching between the branes. A prototype 
for this correspondence and most important in what follows is the
computation of the leading order velocity-dependent potential 
$v^4/r^7$ between two D0-branes in [4,7]. One other interesting 
follo-wup to this correspondence is the realization that at
substringy distances the classical geometry of spacetime is identified
with the quantum moduli space of gauge inequivalent configurations.
See [8], however, for other examples in connection with this phenomenon.
\smallskip

Using these observations along with the description of the 
supermembrane worldvolume action in terms of a supersymmetric 
quantum mechanical system as found in [9] and which was 
reinterpreted later in [10] as being the maximally supersymmetric 
quantum mechincal system describing N D0-branes in the large N 
limit, Banks, Fischler, Shenker and Susskind (BFSS) put forward the more 
far-reaching Matrix theory conjecture [11]. Simply put the
conjecture states that M-theory, in the light-cone frame, 
is exactly described by the large N limit of the 
supersymmetric matrix quantum mechanics of N 
D0-branes\footnote{See however [12] for the new version of 
this conjecture at finite $N$. For the purpose of our work here it 
suffices to deal with the original BFFS Matrix theory of [11].}. 
This conjecture came as an attempt to describe the short-distance 
limit of M-theory. M-theory made its first entry in the string web of 
dualities as the strong coupling limit of type IIA string theory 
[13]. Before the BFFS conjecture, very little 
was known about this theory except that at low energies and large 
distances M-theory is described by the eleven-dimensional supergravity.
The BFSS conjecture (if correct) seems to indicate that all of the
eleven-dimendional physics in the infinite momentum frame is contained
in the maximally supersymmetric gauge theory reduced to the quantum 
mechanics on the worldline of the D0-branes. 
\smallskip

Here is now that come the
relevance of the correspondence between supergravity and SYM theory 
explained above. The exact equivalence (due to supersymmetry) between 
the leading long-distance supergravity interaction $v^4/r^7$ 
between D0-branes governed by a single supergraviton exchange and the 
one-loop matrix theory result becomes an important consistency criteria 
for the BFSS conjecture. In other words for the BFSS conjecture to hold 
it is important that the $v^4/r^7$ potential receives no contrbution 
beyond one-loop on the matrix theory side. The authors of [11] suspected 
the existenece of some non-renormalization theorem in the context of a 
supersymmetric quantum mechanics model with 16 supercharges to protect 
this term from higherloops contribution. This belief is strengthened
by the existence in the litterature of a similar non-renormalization 
theorem for the $F^4_{\mu\nu}$ term in the action of the ten dimensional 
string theory [4, 14].
\smallskip

This question was undertaken first in [15] and later in 
[16] where using the background field method they showed 
that at two-loops the $v^4/r^7$ term is robust. Very recently
in [17] the robustness of this term was completely established 
when the proof of the non-renormalization theorem alluded to above 
was found. The status of the BFFS matrix theory as a candidate for 
the nonperturbative formulation of of M-theory has also 
improved\footnote{See also the work in [18].} after the work of 
the authors in [19] where they ruled out any discrepancy between 
eleven-dimensional supergravity and the BFSS matrix theory for 
three-body scattering as reported earlier in [20].
For a more complete description of the matrix theory and its 
status we refer the reader to the excellent reviews [21].
\smallskip

The BFSS matrix theory describes naturally the ten-dimensional 
IIA superstring theory since the latter is the limit of M-theory 
at weak string couling. In [22] Ishibashi, Kawai, Kitazawa and 
Tsuchiya (IKKT) proposed another matrix model associated with the 
IIB superstring theory, which is in the spirit of the Eguchi-Kawai
[23] large-N reduced ten-dimensional SYM theory. 
This non-perturbative formulation of IIB supertsring theory is called 
the IKKT matrix model. In analogy with the BFSS matrix model where the 
theory is expressed in terms of D0-branes, in the IKKT model 
one expects that the non-perturabtive formulation of type IIB 
superstring theory is described in terms of a
supresymmetric Yang-Mills gauge theory of N D-instantons derived 
upon reduction to a point\footnote{This interpretation is not totally 
exact since the action describing the IKKT matrix model contains an 
extra term that does not arise from any dimensional reduction and 
was tinterpreted in [22] as a kind of a chemical potential.}.
The connection of the IKKT matrix model with the one of BFSS was 
demonstrated in [22] by considering the action of the former on 
some special point of its moduli space of degenerate vacua. 
Contrary to the BFSS case, however, the IKKT matrix model has 
the manifest Lorentz invariance in ten dimensions and so does not 
present us with the awkwardness of the light-cone frame.   
\smallskip

The validity of the IKKT matrix model relies so far on its ability  
to describe the classical D-brane configurations of type IIB superstring 
and their interactions. In [22] a one-loop computation 
in the background of operator-like solutions corresponding 
to a cluster of IIB D-objects with relative motion and 
occupying some region of spacetime has revealed a leading 
$1/r^8$ behavior in the potential between two D-block objects.
This result manifestly agree with the long-range potential obtained in 
the supergravity calculation based on the massless closed string exchange. 
As argued in [22], the $1/r^8$ behavior of the effective 
interaction in the IKKT matrix model ensures the cluster property among 
the D-objects which is important to the N=2 supersymmetry of the IKKT action 
and hence to the dynamical generation of the spacetime coordinates which 
constitutes one of the nicest feature of the IKKT matrix model. 
\smallskip
 
Furthermore and as in the BFSS case, the exact agreement between the  
IKKT matrix theory and the supergravity result is an important
consistency criterion for the IKKT matrix model to describe
type IIB superstring theory. In particular, higher loop effects
on the IKKT matrix side had better not to spoil this correspondence.
So for these reasons the question of the non-renormalization
of the one-loop result becomes also important here. This article 
is an investigation along this line. We start in Section~(2),
with some preliminaries about the IKKT matrix model.
In Section~(3), we deal with the one-loop computation of the effective
action in the general background of multi-D-objetcs with very
large separations from each other. This background is represented by a
block-diagonal operator-like solutions. The one-loop computation is
not new but was considered previously in [21,24] but
we give it here just to fix conventions and introduce notations and 
definitions that set the ground for the two-loop evaluation of the
effective action in Section~(4). Ou basic tool for the two-loop
computation will be the background field methos and we find 
in anolgy with the investigation of in the BFFS case [15] 
that the one-loop result is not renormalized at this order. 
Section~(5) contains some concludings remarks where we argue 
for a possible extension of the non-renormalization theorem 
applicable to the D0-branes [17] to include the IKKT model 
on a point. In the Appendix, we have gathered some technical 
details that arise in the two-loop 
computation of Section~(4).
\smallskip

\section{Some Preliminaries}

The IKKT matrix model is defined by the partition function
\beq
\Z_0 = \sum_{n=1}^{\infty} \int dA_\mu \,d\Phi_\a\, \e^{-S} \,,
\label{aa}
\eeq
which is a second quantized Euclidean field theory, with 
action\footnote{In what follows, we set $\alpha' = 1$ and 
$g_s= 1$.}
\beq
S = {1\over{{g_s}\left(\alpha'\right)^2}}\; 
\left(
-\, {1\over 4}\, Tr\left(\left[A_\mu , A_\nu\right]^2\right) -
{1\over 2}\, Tr\left(
\bar{\Phi}\,  \Gamma^\mu\,
\left[A_\mu ,\Phi\right]\right)
\right) + \beta\, n\, .
\label{ab}
\eeq
Here ${}^{ij}A_{\mu}$ and ${}^{ij}\Phi_{a}$ are
$n \times n$ Hermitian bosonic and fermionic
matrices, respectively\footnote{Our notation follows the
one of reference [22].}. The vector index $\mu$
runs from 0 to 9 and the spinor index $a$ runs from 1 to
32. The fermion $\Phi$ is a Majorana-Weyl spinor which
satisfies the condition $\Gamma_{11}\, \Phi = \Phi$.
\smallskip

The action (\ref{ab}) is invariant under the
$\N = 2$ supersymmetry transformations
\begin{eqnarray}
\delta^{(1)}\, \left({}^{ij}\Phi_a\right) + \delta^{(2)}\, 
\left({}^{ij}\Phi_a\right) & = & {i\over 2}\,
\left({}^{ij}\left[A_\mu , A_\nu\right]\right)\,
\left(\Gamma^{\mu\nu}\,\epsilon\right)_a + \xi_a\, \delta^{ij}\, ,
\nonumber\\
\delta^{(1)}\, \left({}^{ij}A_\mu\right) + \delta^{(2)}\, 
\left({}^{ij}A_\mu\right) & = & i\, \bar{\epsilon}\, 
\Gamma_\mu \, \left({}^{ij}\Phi\right)\, ,
\label{ac}
\end{eqnarray}
where $\epsilon$ and $\xi$ are the supersymmetry parameters, as well 
as under the gauge transformation
\begin{eqnarray}
\delta_{gauge}\, A_\mu & = & i\, \left[A_\mu\, ,\, \omega\right]\, ,
\nonumber\\
\delta_{gauge}\, \Phi_a & = &  i\, \left[\Phi_a\, ,\, \omega\right]\, .
\label{ad}
\end{eqnarray}
\smallskip

The fromulas (\ref{ac}) and (\ref{ad}) look like as if 
10D SYM theory is reduced to a point. For instance
all the spacetime derivatives drop out from the
non-Abelian field strength $F_{\mu\nu} = i\, [ A_\mu , A_\nu ]$.
However, the action (\ref{ab}) coincides with the one
of 10D SYM theory in the zero volume limit only if $\beta =$ and
$n$ fixed. In [22], the $\beta$ parameter was interpreted as a
kind of chemical potential generated form the 
one-loop renormalization of the
action (\ref{ab}) even if initially $\beta$ is set to zero.
\smallskip

So the action (\ref{ab}) is, up to the $\beta$-term, the low-energy
effective action of a D-instanton of charge $n$ [2].
The other higher dimensional branes are represented by the solutions of the
calssical equations of motion
\beq
\left[A_\mu , \left[A_\mu , A_\nu\right]\right] = 0\,,\quad
\left[A_\mu , \left(\Gamma^\mu\, \Phi\right)_a\right] = 0\, ,
\label{ae}
\eeq
which are to be solved by $n \times n$ matrices $A_\mu$ at infinite $n$.
A general solution has a block-diagonal form
\beq
A_\mu = \left(\begin{array}{cccc}
{}^1y_\mu & & &\\
 &  {}^2y_\mu & & \\
& &  {}^3y_\mu &  \\
& & &  \ddots
\end{array}
\right)\, , \quad \Phi_a = 0\, ,
\label{af}
\eeq
where ${}^iy_\mu$ $(i=1,2,3,\dots)$ is a non-diagonal $n_i \times n_i$
matrix. We may regard ${}^iy_\mu$ as a D-object occupying
some region of spacetime. A much simpler solution corresponds
to the case where the ${}^iy_\mu$'s become diagonal matrices. 
Using this fact, we can decompose ${}^iy_\mu$ in the general
solution (\ref{af}) as
\begin{eqnarray}
^iy_\mu & = & {}^id_\mu\, 1_{n_i} + {}^ip_\mu\, ,
\nonumber\\
Tr\, {}^ip_\mu & = & 0 \, ,
\label{ag}
\end{eqnarray}
where ${}^id_\mu$ is a real number representing the center of mass coordinate
of the $i$-th block. More on the notations we are adopting for
the block elements ${}^iy_\mu$, ${}ip_\mu$ and ${}^id_\mu$ in the next section.
For the purpose of our work here, we assume through out 
all the paper that the blocks are separated far enough from each other so that
for all $i$ and $j$'s, $({}^id_\mu - {}^jd_\mu)^2$ are large. 
\smallskip

It was argued in [22,25] that for the classical 
solutions representing BPS states the field strength should be
proportional to the unit matrix, that is, 
\beq
{}^if_{\mu\nu} = i\, \left[{}^ip_\mu\, ,\, {}^ip_\nu\right] 
= c_{\mu\nu}\, 1_{n_i}\, ,
\label{ah}
\eeq
or since $f_{\mu\nu}$ has a block-diagonal form we can also write
\beq
f_{\mu\nu} = \left(\begin{array}{cccc}
{}^1f_{\mu\nu} & & &\\
& {}^2f_{\mu\nu} & &\\
& & {}^3f_{\mu\nu} &\\
& & & \ddots
\end{array}\right)\, .
\label{ahh}
\eeq
The classical equations (\ref{ae}) are in this case automatically
satisfied. Since D-branes are BPS states [1], the classical
solutions of the matrix model which correspond to D-branes should 
have this property.
\smallskip

\section{The One-Loop Effective Action}

In this section, we calculate the one-loop effective action for 
the interaction between many diagonal blocks. The calculation is 
similar to the one performed in [22] and we repeat here only to fix
notation and prepare the ground for the derivation of the two-loop
effective action in Section~(4). Using the background field method,
we decompose the matrices $A_\mu$ and $\Phi$ into a general background
having a block-diagonal form plus the quantum fluctuations. Namely,
\begin{eqnarray}
{}^{ij}A_\mu^{qr} & = & {}^id_\mu\,\delta^{ij}\,\delta^{qr}
+ {}^ip_\mu^{qr}\, \delta^{ij} + {}^{ij}a_\mu^{qr}\, ,
\label{ba}\\
{}^{ij}\Phi_a^{qr} & = & {}^{ij}\phi_a^{qr} + {}^{ij}\varphi_a^{qr}\, ,
\label{bb}
\end{eqnarray}
where the matrix elements ${}^{ij}a_\mu^{qr}$ and ${}^{ij}\varphi_a^{qr}$ 
are the bosonic and the fermionic quantum fluctuations, respectively.
\smallskip

Since these quantum fluctuations will come often in this paper, a word on 
the notation we are adopting for them is in order. This proves also useful 
later when we come to the derivation of the Feynman rules involved in the 
two-loop computation. The quantum fields matrices $a_\mu$ and $\varphi_a$ belong 
to the general space of matrices. A general matrice $X$ on this space can be 
denoted by its $\left(i, j\right)$ blocks as an $n_i \times n_j$ matrix 
${}^{ij}X$. So the block elements 
of a block-diagonal matrix $X$ satisfy ${}^{ij}X = {}^{i}X\,\delta^{ij}$,
where ${}^{i}X$ is an $n_i \times n_i$ matrix. The matrix elements of 
a block element matrix ${}^{ij}X$ are given by ${}^{ij}X^{qr}$. For instance
the matrix $d_\mu$ representing the center of mass coordinates 
of some $i$-th block has a block diagonal form and we denote its elements
as ${}^{ij}d_\mu^{qr} = 
{}^{i}d_\mu\,\delta^{ij}\,\delta^{qr}$, where ${}^{i}d_\mu$
is a pure real number.
\smallskip

Now, we move to the derivation of the one-loop effective action. 
As is usual in a gauge theory (which is the case here) one should 
add both the gauge fixing and the ghost terms to the action (\ref{ab}), 
such terms are given by
\beq
S_{gf} = -\, {1\over 2}\, Tr\, \left(\left[d_\mu + p_\mu\, ,\, 
a_\mu\right]^2\right) - Tr\, \left(\left[d_\mu + p_\mu\, ,
\, \bar{b}\right]\,
\left[d_\mu + p_\mu\, ,\, c\right]\right)\, ,
\label{bc}
\eeq
where the matrices $c$ and $\bar{b}$ represent the ghosts and 
anit-ghosts, respectively. In the following, adding $S_{gf}$ to
the action (\ref{ab}) expanded in the quantum fluctuations
$a_\mu$ and $\varphi_a$ and using the classical equations
of motion (\ref{ae}) after setting the fermionic background
$\phi$ to zero, we find
\beq
S + S_{gf} = S_2 + S_4^B + S_3^B + S_3^F + S_3^{{ghost}}\, .
\label{bd}
\eeq
The action $S_2$ is obtained by keeping the quantum fluctuations
up to second order such as
\begin{eqnarray}
S_2 & = &  {1\over 2}\, Tr\, \left(a_\mu\, \left[p_\mu\, ,\,
\left[p_\mu\, ,\, a_\nu\right]\right]\right)
+ Tr\, \left(a_\mu\, \left[\left[p_\mu\,,\, p_\nu\right]\, , 
\, a_\nu\right]\right)
- {1\over 2}\, Tr\, \left(\bar{\varphi}\, \Gamma^{\mu}\, 
\left[p_\mu\, ,\, \varphi\right]\right)
\nonumber\\
& &
\, +\,  Tr\, \left(\bar{b}\,\left[p_\mu\, ,\, \left[p_\mu\, ,
\, c\right]\right]\right)\, .
\label{be}
\end{eqnarray}
$S_4^B$ involves the quartic interactions among the bosonic 
quantum fluctuations $a_\mu$ and $S_3^B$ the cubic ones
\begin{eqnarray}
S_4^B & = & -\, {1\over 4}\, Tr\, \left(\left[a_\mu\, ,
\, a_\nu\right]^2\right) 
\nonumber\\
& = & A_{\mu\nu\lambda\rho}\; Tr\,\left(a_\mu\, a_\nu\, 
a_\lambda\, a_\rho\right)\, ,
\label{bf}\\
S_3^B & = & -\, Tr\, \left(\left[d_\mu + 
p_\mu\, ,\, a_\nu\right]\,
\left[a_\mu\, ,\, a_\nu\right]\right)
\nonumber\\
& = & B_{\mu\nu\lambda\rho}\; 
Tr\,\left[\left(d_\mu + p_\mu\right)\, 
a_\nu\, a_\lambda\, a_\rho\right]\, ,
\label{bg}
\end{eqnarray}
where $A_{\mu\nu\lambda\rho}$ and $B_{\mu\nu\lambda\rho}$ 
are given by
\begin{eqnarray}
A_{\mu\nu\lambda\rho} & = & {1\over 2}\, \left(\delta_{\mu\nu}\, 
\delta_{\lambda\rho}
\, -\, \delta_{\mu\lambda}\, \delta_{\nu\rho}\right)\, ,
\label{bh}\\
B_{\mu\nu\lambda\rho} & = & \left(\delta_{\mu\nu}\, \delta_{\lambda\rho}\,
+\,
\delta_{\mu\rho}\, \delta_{\nu\lambda}\, -\, 2\, \delta_{\mu\lambda}\, 
\delta_{\nu\rho}\right)\, .
\label{bi}
\end{eqnarray}
$S_3^F$ is the action involving cubic interactions of both 
fermionic $\varphi$ and bosonic $a_\mu$ fluctuations and is
given by
\beq
S_3^F = -\, {1\over 2}\, Tr\, \left(\bar{\varphi}\, \Gamma^{\mu}\, 
\left[a_\mu\, ,\, \varphi\right]\right)\, .
\label{bj}
\eeq
Similarly $S_3^{{ghost}}$ includes the cubic ghost interactions
\beq
S_3^{{ghost}} = -\, Tr\, \left(\left[p_\mu\, ,\, \bar{b}\right]\,
\left[a_\mu\, ,\, c\right]\right)\, .
\label{bk}
\eeq
\smallskip

In order to compute the one-loop effective action, it suffices to consider
the quantum fluctuations in the action only up to the second order. The action
$S_2$ contains all such terms. To perform the functional integration involved
in the evaluation of the one-loop effective action 
\beq
\W^{(1)} = -\, \log \int\,
\D a_\mu\, \D \varphi_a\, \D c\, \D \bar{b}\; e^{-\, S_2}\, ,
\label{bl}
\eeq
it will be convenient to introduce the notation in which the
the adjoint operators $P_\mu$ and $F_{\mu\nu}$ act on the space 
of matrices as 
\begin{eqnarray}
P_\mu\, X & = & \left[d_\mu + p_\mu + a_\mu\, ,\, X\right]\, ,
\label{bm}\\
F_{\mu\nu}\, X & = & \left[f_{\mu\nu}\, ,\, X\right] =
\left[i\,\left[p_\mu\, ,\, p_\nu\right]\, ,\, X\right]\, .
\label{bn}
\end{eqnarray} 
Using this notation, we can write the action $S_2$ as
\beq
S_2 = {1\over 2}\, Tr\, \left[a_\mu\, 
\left(P^2\, \delta_{\mu\nu} - 2\, i\, F_{\mu\nu}\right)\, a_\nu\right]
- {1\over 2}\, Tr\,\left(\bar{\varphi}\, \Gamma^\mu \, P_\mu\, 
\varphi\right) + Tr\,\left(\bar{b}\, P^2\, c\right)\, .
\label{bo}
\eeq
From this the one-loop effective action $\W^{(1)}$ in 
(\ref{bl}) is evaluated easily and is given by
\begin{eqnarray}
\W^{(1)} & = & {1\over 2}\, Tr\, \left(P^2\, \delta_{\mu\nu}\, -\, 
2\, i\, F_{\mu\nu}\right)\, -\, 
{1\over 4}\, Tr\, \left[\left(P^2\, +\, {i\over 2}\, F_{\mu\nu}\, 
\Gamma^{\mu\nu}\right)\,
\left({{1 + \Gamma_{11}}\over{2}}\right)\right]
\nonumber\\
& & \, -\, Tr\,\left(P^2\right)\, .
\label{bp}
\end{eqnarray}
The cases where 
${}^{ij}f_{\mu\nu}^{qr} = c_{\mu\nu}\, \delta^{ij}\, \delta^{qr}$ 
have special meaning. These correspond to BPS-saturated states
backgrounds [22,24,26]. Since $F_{\mu\nu} = 0$ in these cases, we have
\beq
\W^{(1)} = \left({1\over 2}\, .\,  10 \, -\, {1\over 4}\, .\, 16
\, -\, 1\right)\, Tr\, \left(P^2\right) = 0\, ,
\label{bq}
\eeq
which means that the one-loop quantum corrections vanish due to
supersymmetry. This is consistent with the well known fact that
the BPS-saturated states have no quantum corrections which also
ensures their stability. The simplest example is the BPS
plane vacuum for which the background matrix elements are
given by ${}^{ij}p_\mu^{qr} = {}^{i}d_\mu\,\delta^{ij}\,\delta^{qr}$
and in this case $f_{\mu\nu} = 0$ trivially.
\smallskip

For a general non-BPS background such as $F_{\mu\nu} \neq 0$, 
the one-loop effective action (\ref{bp}) expanded in the 
inverse power of $\left({}^id_\mu - {}^jd_\mu\right)^2$ is 
given by
\begin{eqnarray}
\W^{(1)} & = & -\, Tr\,\left({1\over P^2}\, F_{\mu\nu}\, 
{1\over P^2}\, F_{\nu\lambda}\, 
{1\over P^2}\, F_{\lambda\rho}\, {1\over P^2}\, F_{\rho\mu}\right)
\nonumber\\
& &\, -\, 2\, 
Tr\,\left({1\over P^2}\, F_{\mu\nu}\, {1\over P^2}\, F_{\nu\lambda}\, 
{1\over P^2}\, F_{\mu\rho}\, {1\over P^2}\, F_{\rho\lambda}\right)
\nonumber\\
& & +\, {1\over 2}\, 
Tr\,\left({1\over P^2}\, F_{\mu\nu}\, {1\over P^2}\, F_{\mu\nu}\, 
{1\over P^2}\, F_{\lambda\rho}\, {1\over P^2}\, F_{\lambda\rho}\right)\, 
\nonumber\\
& &+\, {1\over 4}\,
Tr\,\left({1\over P^2}\, F_{\mu\nu}\, {1\over P^2}\, F_{\lambda\rho}\, 
{1\over P^2}\, F_{\mu\nu}\, {1\over P^2}\, F_{\lambda\rho}\right)
\nonumber\\
& & \, +\, \O\left(F^5\right)\, ,
\label{br}
\end{eqnarray}
where we have used the following identities for the Dirac
matrices
\begin{eqnarray}
\{\Gamma^\mu\, ,\, \Gamma^\nu\} & = &  -\, 2\, 
\delta^{\mu\nu}\,,\quad 
\Gamma^{\mu\nu} = {1\over 2}\, \left[\Gamma^\mu\, ,\, 
\Gamma^\nu\right]\,,\quad
Tr\, \left(\Gamma^\mu\, \Gamma^\nu\right) =  -\, 32\, 
\delta^{\mu\nu}\, ,
\nonumber\\ 
Tr\,\left(\Gamma^\mu\, \Gamma^\nu\, \Gamma^\lambda\, 
\Gamma^\rho\right) & =  &
32\, \left(
\delta^{\mu\nu}\, \delta^{\lambda\rho} -
\delta^{\mu\lambda}\, \delta^{\nu\rho} +
\delta^{\mu\rho}\, \delta^{\nu\lambda}\right)\, .
\label{brx}
\end{eqnarray} 
Although this background is not BPS-saturated there is still some
left assymptotic residual $N=2$ supersymmetry which ensures the 
cancellation between the bosonic and the fermionic contribution up
to the third ordr in $F_{\mu\nu}$. This feature is very reminiscent 
of the results of many previous investigations 
[3,4,14] with $N=2$ supersymmetry. It is not surprising that we recover here 
the same kind of cancellation since as argued originally in [22] 
the IKKT matrix model is $T$-dual to the BFFS matrix 
model where this phenomenon appear also.
In fact one of the goals of Section~(4) is to show that this cancellation 
continues to hold even at two-loops.
\smallskip

To simplify further the expression of (\ref{br}), we need to introduce 
more notations such as
\begin{eqnarray}
{}^{ij}\left(P_\mu\, X\right) & = & 
{}^{ij}d_\mu\, {}^{ij}X\, +\, {}^{ij}p_\mu\, {}^{ij}X \,
\equiv {}^{ij}P_\mu\, {}^{ij}X\, ,
\nonumber\\
{}^{ij}d_\mu\, {}^{ij}X & = & 
\left({}^id_\mu - {}^jd_\mu\right)\, {}^{ij}X\, , \quad 
\left({}^id_\mu - {}^jd_\mu\right)
\hbox{act as real numbers}\, ,
\nonumber\\
{}^{ij}p_\mu\, {}^{ij}X & = &
{}^ip_\mu\, {}^{ij}X\, -\, {}^{ij}X\, {}^{j}p_\mu\, .
\label{bs}
\end{eqnarray}
Using these definitions, it is easy to show that
\beq
{}^{ij}\left(P^2\, X\right) = {}^{ij}\left(P^2\right)
\, {}^{ij}X = \left({}^{ij}d^2\right)\, {}^{ij}X + 2\, 
{}^{ij}d\, .\, {}^{ij}p\, {}^{ij}X 
+ \left({}^{ij}p^2\right)\, {}^{ij}X\, ,
\label{bss}
\eeq
where
\begin{eqnarray}
\left({}^{ij}p^2\right)\, {}^{ij}X & = &
\left({}^ip_\mu^2\right)\, {}^{ij}X 
+ {}^{ij}X\, \left({}^jp_\mu^2\right) 
- 2\, {}^ip_\mu\, {}^{ij}X\, {}^jp_\mu\, ,
\nonumber\\
{}^{ij}d\, .\, {}^{ij}p\, {}^{ij}X & = &
\left({}^id_\mu - {}^jd_\mu\right)\,
\left({}^ip_\mu\, {}^{ij}X - {}^{ij}X\, {}^jp_\mu\right)\, ,
\nonumber\\
\left({}^{ij}d^2\right)\, {}^{ij}X & = &
\left({}^id_\mu - {}^jd_\mu\right)^2\, {}^{ij}X\, .
\label{bsss}
\end{eqnarray}
In the same way, we can decompose $F_{\mu\nu}$ as
\begin{eqnarray}
{}^{ij}\left(F_{\mu\nu}\, X\right) & = &
{}^{ij}f_{\mu\nu}\, {}^{ij}X \equiv {}^{ij}F_{\mu\nu}\, {}^{ij}X\, ,
\nonumber\\
{}^{ij}f_{\mu\nu}\, {}^{ij}X & = &
{}^if_{\mu\nu}\, {}^{ij}X\, -\, {}^{ij}X\, {}^jf_{\mu\nu}\, .
\label{bt}
\end{eqnarray}
It is clear from the above expressions that not only does $P_\mu$ and
$F_{\mu\nu}$ operate on each block ${}^{ij}X$ independently
but also their action on the left or on the right of ${}^{ij}X$
are totally independent. It follows then that the trace of an operator
$O$ such as the one appearing in $\W^{(1)}$ in (\ref{br}) and which 
consists of $P_\mu$ and $F_{\mu\nu}$ is evaluated using this fromula
\beq
Tr\, \left(O\right) = \sum_{i,j = 1}^{n}\; 
Tr\, {}^{ij}O_L\;
Tr\, {}^{ij}O_R\, .
\label{bu}
\eeq
Using these definitions we can easily evaluate the one-loop effective 
action $\W^{(1)}$ and the result is a sum of contributions from each  
$\left(i\, ,\, j\right)$ block $\W^{\left(i\, ,\, j\right)}$ which
describes the effective interactions between the $i$-th and $j$-th 
blocks, that is, 
\beq
\W^{(1)} = \sum_{i,j = 1}^{n}\;\,\W^{\left(i\, ,\, j\right)}\, ,
\label{bv}
\eeq
where $\W^{\left(i\, ,\, j\right)}$ are given to leading order
for large separation $({}^id - {}^jd)^2$ by
\begin{eqnarray}
\W^{\left(i\, ,\, j\right)} & = &
{1\over{4\, \left({}^{ij}d\right)^8}}\, 
\left[
-\, 4\, n_j\, Tr\,\left({}^if_{\mu\nu}\, {}^if_{\nu\lambda}\, 
{}^if_{\lambda\rho}\, {}^if_{\rho\mu}\right)
\, - 8\, n_j\, Tr\,\left({}^if_{\mu\nu}\, {}^if_{\nu\lambda}\, 
{}^if_{\mu\rho}\, {}^if_{\rho\lambda}\right)\right.
\nonumber\\
& & +\,2\, n_j\, Tr\,\left({}^if_{\mu\nu}\, {}^if_{\mu\nu}\, 
{}^if_{\lambda\rho}\, {}^if_{\lambda\rho}\right)
\, +\, n_j\, Tr\,\left({}^if_{\mu\nu}\, {}^if_{\lambda\rho}\, 
{}^if_{\mu\nu}\, {}^if_{\lambda\rho}\right)
\nonumber\\
& & -\, 4\, n_i\, Tr\,\left({}^jf_{\mu\nu}\, {}^jf_{\nu\lambda}\, 
{}^jf_{\lambda\rho}\, {}^jf_{\rho\mu}\right)
\, - 8\, n_i\, Tr\,\left({}^jf_{\mu\nu}\, {}^jf_{\nu\lambda}\, 
{}^jf_{\mu\rho}\, {}^jf_{\rho\lambda}\right)
\nonumber\\
& & +\,2\, n_i\, Tr\,\left({}^jf_{\mu\nu}\, {}^jf_{\mu\nu}\, 
{}^jf_{\lambda\rho}\, {}^jf_{\lambda\rho}\right)
\, +\, n_i\, Tr\,\left({}^jf_{\mu\nu}\, {}^jf_{\lambda\rho}\, 
{}^jf_{\mu\nu}\, {}^jf_{\lambda\rho}\right)
\nonumber\\
& &\left.
-\, 48\, Tr\,\left({}^if_{\mu\nu}\, {}^if_{\nu\lambda}\right)\;
Tr\,\left({}^jf_{\mu\rho}\, {}^jf_{\rho\lambda}\right)
\, +\, 6\, Tr\,\left({}^if_{\mu\nu}\, {}^if_{\mu\nu}\right)\;
Tr\,\left({}^jf_{\lambda\rho}\, {}^jf_{\lambda\rho}\right)
\right]
\nonumber\\
& & +\, \O\left(1/\left( {}^{ij}d\right)^9\right)\, .
\label{bw}
\end{eqnarray}
In [22] the last two terms were identified with the exchange
of the graviton and the scalar dilaton. After this review, 
we are now ready to move to the two-loop computation.
\smallskip

\section{The Two-Loop Effective Action}

In this section we shall carry out the derivation
of the two-loop effetcive action of the IKKT matrix gauge
system described by the total action in (\ref{bd}).
A similar calculation was performed for the BFFS
matrix model in [15,16,19] using the
background field method. Following the
same approach [27], we treat the background field 
${}^id_\mu + {}^ip_\mu$ in (\ref{ba}) exactly so that
it enters in the propagators and vertices of the 
theory. Therefore to compute
the gauge invariant background field effective action at two-loops
one has to sum only over all 1PI vacumm diagrams (without external lines) 
involving quartic and cubic vertices as dictated by the action (\ref{bd}).
The two-loop vacumm graphs in question are displayed in Fig(1) and are
adapted for our purposes to describe the IKKT matrix model [22].
We shall return back to explaining our representation of these 
graphs in Fig~(1) after introducing below the propagators for the bosonic,
fermionic and ghost fields.
\smallskip

Knowing that at two-loops the effective action is given by the 
four vacuum graphs in Fig~(1) allows us to represent it as a sum of 
conctributions arising from each of the interactions $S_4^B$, $S_3^B$, 
$S_3^F$ and $S_3^{{ghost}}$, that is,
\beq
\W^{(2)} = \W_4^B + \W_3^B + \W_3^F + \W_3^{{ghost}}\, ,
\label{ca}
\eeq
where $\W_4^B$ and $\W_3^B$ account for the grahps in Fig~(1) involving 
the quartic and the cubic vertices involving the bosonic fluctuations
$a_{\mu}$. $\W_3^F$ accounts for the cubic vertex graph involving
the fermionic fluctuations $\varphi_a$ and $\W_3^{ghost}$ describe the 
cubic vertex graph with ghost fields. The computation of $\W^(2)$ will boil 
down then to evaluating the terms $\W_4^B$, $\W_3^B$, $\W_3^F$, 
$\W_3^{{ghost}}$ individually. For this we need to know the Green's 
functions of the theory under consideration. It is a well known 
result of quantum field theory that the effective action is expressed
as a product of the Green's functions of the theory with the
appropriate `contractions'. Next we shall describe such Green's
functions. 
\smallskip

\subsection{Feynman Rules and Diagrammatics}

The Green's functions of interest are determined by taking
the functional derivatives with respect to the different
bosonic and fermionic source functions in the whole
generating functional of the IKKT matrix system (\ref{bl})
considered after adding to $S_2$ the interacting part
$S_{{int}}$ given by
\beq
S_{{int}} = S_4^B + S_3^B + S_3^F + S_3^{{ghost}}\, . 
\label{cab}
\eeq
To escape unnecessary details we display below only the key formulas 
used to evaluate the different terms in the two-loop effective 
action (\ref{ca}). After adding the source functions $J_\mu$, 
$\bar{\eta}$, $\eta$, $\bar{\psi}$ and $\chi$ the complete free 
generating functional is given as a product
\beq
\Z_0\left[J , \bar{\eta} , \eta , \bar{\psi} , \chi\right] =
\Z_0^B\left[J\right]\; \Z_0^F\left[\bar{\eta} , \eta\right]\;
\Z_0^{{ghost}}\left[\bar{\psi} , \chi\right]\, ,
\label{cb}
\eeq
where
\begin{eqnarray}
\Z_0^B\left[J\right] & = & \int\; \D a_\mu\; e^{-\, {1\over 2}\, 
Tr\, \left(a_\mu . \A_{\mu\nu} . a_\nu\right) +
Tr\, \left(J_\mu . a_\mu\right)} = e^{{1\over 2}\, 
Tr\, \left(J_\mu . G_{\mu\nu} . J_\nu\right)}\, ,
\label{cc}\\
\Z_0^F\left[\bar{\eta} , \eta\right] & = &
\int\; \D \bar{\eta}\, \D \eta\; 
e^{{1\over 2}\, Tr\, \left(\bar{\varphi} . \S . \varphi\right)
+ Tr\, \left(\bar{\varphi} . \eta\right) + 
Tr\, \left(\bar{\eta} . \varphi\right)} = e^{-\, 2\, 
Tr\, \left(\bar{\eta} . H . \eta\right)}\, ,
\label{cd}\\
\Z_0^{{ghost}}\left[\bar{\psi} , \chi\right] & = &
\int\; \D \bar{b}\, \D c\;
e^{-\, Tr\, \left(\bar{b} . \T . c\right) +
Tr\, \left(\bar{b} . \chi\right) +
tr\, \left(\bar{\psi} . c\right)} =
e^{Tr\, \left(\bar{\psi} . E . \chi\right)}\, ,
\label{ce}
\end{eqnarray}
and
\beq
\A_{\mu\nu} = P^2\, \delta_{\mu\nu} - 2\, i\, F_{\mu\nu}\, ,\quad
\S = \Gamma^\mu\, P_\mu\, ,\quad
\T = P^2\, .
\label{cf}
\eeq
\smallskip

In the functional approach, the Green's functions are given by the
inverse of the operators appearing in the quadratic part of the
action. From the formulas above (\ref{cc}), (\ref{cd}) and (\ref{ce}) 
we can straightforwardly read the Green's functions of our theory.
For the bosonic quantum flcutuation we have
\begin{eqnarray}
G_{\mu\nu} & = & \A^{-1}_{\mu\nu} = \left(P^2\, \delta_{\mu\nu} - 
2\, i\, F_{\mu\nu}\right)^{-1}\, ,
\nonumber\\
& = & {1\over P^2}\, \delta_{\mu\nu} + 2\, i\, {1\over P^2}\, 
{1\over P^2}\, F_{\mu\nu}
- 4\, {1\over P^2}\, {1\over P^2}\, F_{\mu\alpha}\, 
{1\over P^2}\, F_{\alpha\nu}
+ \O\left(F^3\, ,\, 1/P^8\right)\, ,
\label{cg}
\end{eqnarray}
where we have expanded the propagator in the inverse power of $1/P^2$ 
since eventually we are interested (as in the one-loop) only in the long 
range contributions of the vacuum graphs emerging at two-loops. 
Such interactions will in principle add up to the one-loop result in 
(\ref{br}) and (\ref{bw}) and constitute higher order corrections. 
The fermionic quantum fluctuations on the other hand involves the 
Green's function
\begin{eqnarray}
H & = & \S^{-1} = \left(P\!\!\!\!/\right)^{-1} = 
\left(\Gamma^\mu\, P_\mu\right)^{-1} = -\, P\!\!\!\!/\, 
{1\over P^2}\, 
\left(1 + {i\over 2}\, \Gamma^{\mu\nu}\, {1\over P^2}\, 
F_{\mu\nu}\right)^{-1}\, ,
\nonumber\\
& = & -\, P\!\!\!\!/\, {1\over P^2}\, + {i\over 2}
P\!\!\!\!/\,\Gamma^{\mu\nu}\, 
{1\over P^2}\, {1\over P^2}\, F_{\mu\nu} +
{1\over 4}\, P\!\!\!\!/\, \Gamma^{\mu\nu}\, \Gamma^{\alpha\beta}\,
{1\over P^2}\, {1\over P^2}\, F_{\mu\nu}\, {1\over P^2} 
F_{\alpha\beta}
+ \O\left(F^3 , 1/P^7\right).
\label{ch}
\end{eqnarray}
For the ghost fields the Green's function is simply
\beq
E = \T^{-1} = {1\over P^2}\, .
\label{ci}
\eeq
\smallskip

In the above formulas what we mean by our notation
$\Tr\, \left(J_\mu . G_{\mu\nu} . J_\nu\right)$ is the
following
\begin{eqnarray}
\Tr\, \left(J_\mu . G_{\mu\nu} . J_\nu\right) & = &
\sum_{i,j = 1}^{n}\; \sum_{p=1}^{n_i}\;\sum_{q=1}^{n_j}\;
\sum_{r=1}^{n_j}\;\sum_{s=1}^{n_i}\;
\left({}^{ij}J_\mu^{pq}\right)\;
\left({}^{ji}G_{\mu\nu}{{}^{qr}_{sp}}\right)\;
\left({}^{ji}J_\mu^{rs}\right)\, ,
\nonumber\\
& \equiv & \sum_{i,j}\; \sum_{p,q,r,s}\; 
\left({}^{ij}J_\mu^{pq}\right)\;
\left({}^{ji}G_{\mu\nu}{{}^{qr}_{sp}}\right)\;
\left({}^{ji}J_\nu^{rs}\right)\, ,
\label{cfa}
\end{eqnarray}
where we have used our observation of Section~(3) 
which indicates that an operator such as 
$G_{\mu\nu}$ consisting of only $P_\mu$ and $F_{\mu\nu}$
(since it is the inverse of $\A_{\mu\nu}$)
must act on the blocks sources ${}^{ij}J_\mu$ independently. 
Moreover it should act both on the left and on the right of 
${}^{ij}J_\mu$ according to the rules (\ref{bs}),
(\ref{bss}), (\ref{bsss}) and (\ref{bt}) derived in 
Section~(3). The different sums in (\ref{cfa}) are easy to 
understand recalling that the block source ${}^{ij}J_\mu$ 
is a $n_i \times n_j$ block matrix and that the order and the 
range over which the indices $p,q,r,s$ are summed is in such 
a way to respect the usual matrix product and the trace cyclic 
property. Note also that the block indices $(i,j)$ are arranged 
so as to repsect the matrix product and the trace formula. 
The sum over the repeated spacetime indices $\mu ,\nu$ is also 
understood. Applying the same remarks above to the fermionic 
and ghost terms, we have readily
\begin{eqnarray}
Tr\, \left(\bar{\eta} . H . \eta\right) & = &
\sum_{i,j = 1}^{n}\; \sum_{p=1}^{n_i}\;\sum_{q=1}^{n_j}\;
\sum_{r=1}^{n_j}\;\sum_{s=1}^{n_i}\;
\left({}^{ij}\bar{\eta_a}^{pq}\right)\; 
\left({}^{ji}H_{ab}{{}^{qr}_{sp}}\right)\;
\left({}^{ji}{\eta_b}^{rs}\right)\, ,
\nonumber\\
& \equiv &
\sum_{i,j}\; \sum_{p,q,r,s}\; 
\left({}^{ij}\bar{\eta_a}^{pq}\right)\; 
\left({}^{ji}\H_{ab}{{}^{qr}_{sp}}\right)\;
\left({}^{ji}{\eta_b}^{rs}\right)\, ,
\label{cfb}
\end{eqnarray}
where the sum over the repeated Dirac indices $a,b$ above is 
understood. For the ghosts we have
\begin{eqnarray}
Tr\, \left(\bar{\psi} . E . \chi\right) & = &
\sum_{i,j = 1}^{n}\; \sum_{p=1}^{n_i}\;\sum_{q=1}^{n_j}\;
\sum_{r=1}^{n_j}\;\sum_{s=1}^{n_i}\;
\left({}^{ij}\bar{\psi}^{pq}\right)\; 
\left({}^{ji}E{{}^{qr}_{sp}}\right)\;
\left({}^{ji}\chi^{rs}\right)\, ,
\nonumber\\
& \equiv &
\sum_{i,j}\; \sum_{p,q,r,s}\;
\left({}^{ij}\bar{\psi}^{pq}\right)\; 
\left({}^{ji}E{{}^{qr}_{sp}}\right)\;
\left({}^{ji}\chi^{rs}\right)\, .
 \label{cfc}
\end{eqnarray}
Using these properties, we shall illustrate in 
the Appendix using some examples that arises in the computation 
of $\W_4^B$, $\W_3^B$, $\W_3^F$ and $\W_3^{{ghost}}$ below how in
the trace operation we take into account the fact that the operators 
${}^{ij}G_{\mu\nu}{{}^{qr}_{sp}}$, ${}^{ij}H_{ab}{{}^{qr}_{sp}}$ 
and ${}^{ij}E{{}^{qr}_{sp}}$ act on both the left and the right 
of the block matrices. Finally, it is clear that the above remarks
carry over similarly to our notation of the quadratic terms
$Tr\, \left(a_\mu . \A_{\mu\nu} . a_\nu\right)$,
$Tr\, \left(\bar{\varphi} . \S . \varphi\right)$ and 
$Tr\, \left(\bar{b} . \T . c\right)$
in (\ref{cb}), (\ref{cd}) and (\ref{ce}).
\smallskip

We are now in a position to derive the explicit expressions of each 
of the two-loop vacuum graphs in Fig~(1). By taking into account the 
matrix nature of the quantum flcutuations on which the propgators act 
we have indicated bosonic propagator ${}^{ij}G_{\mu\nu}{{}^{pq}_{rs}}$ 
by two wavy lines where we put on each the appropriate indices as 
dictated by the formulas in (\ref{cfa}). The fermionic propagator 
${}^{ij}H_{ab}{{}^{pq}_{rs}}$ is indicated by two solid lines with the 
indices as in (\ref{cfb}) and the ghost propagator ${}^{ij}E{{}^{pq}_{rs}}$
by two dashed lines and with the incies as in (\ref{cfc}). To calculate
$\W^{(2)}$, we proceed perturbatively treating exp$(-\, S_{{int}})$
as a power series in the formula
\beq
\Z\left[J,\bar{\eta},\eta,\bar{\psi},\chi\right] =
e^{-\, S_{{int}}
\left(
{\delta\over{\delta J}}\, ,
{\delta\over{\delta \bar{\eta}}}\, , {\delta\over{\delta \eta}}\, ,
{\delta\over{\delta \bar{\psi}}}\, , {\delta\over{\delta \chi}}
\right)}\; 
\Z_0^B\left[J\right]\; \Z_0^F\left[\bar{\eta} , \eta\right]\;
\Z_0^{{ghost}}\left[\bar{\psi} , \chi\right]\, ,
\label{cj}
\eeq
where we have make the following substitutions in 
$S_4^B$, $S_3^B$, $S_3^F$ and $S_3^{{ghost}}$ 
\begin{eqnarray}
{\delta}\over{\delta\, {}^{ij}J_\mu^{pq}}
& \longrightarrow & 
{}^{ji}a_\mu^{qp}\, ,
\label{cka}\\
{\delta}\over{\delta\, {}^{ij}\eta_a^{pq}} 
& \longrightarrow & 
-\; {}^{ji}\bar{\varphi_a}^{qp}\, ,
\label{ckb}\\
{\delta}\over{\delta\, {}^{ij}\bar{\eta_a}^{pq}} 
& \longrightarrow & 
{}^{ji}{\varphi_a}^{qp}\, ,
\label{ckc}\\
{\delta}\over{\delta\, {}^{ij}\chi^{pq}} 
& \longrightarrow & 
-\; {}^{ji}\bar{b}^{qp}\, ,
\label{ckd}\\
{\delta}\over{\delta\, {}^{ij}\bar{\psi}^{pq}} 
& \longrightarrow & 
{}^{ji}c^{qp}\ .
\label{cke}
\end{eqnarray}
Escaping further details since at this point our treatment 
becomes very similar to the usual steps encountered in 
perturbative field theory [28], we can evaluate
the different terms constituting the two-loop effective action
$\W^{(2)}$ in (\ref{ca}). We find that the quartic and cubic bosonic 
interactions contribute respectively the following terms
\beq
\W_4^B = -\; {1\over 2}\;\; B_{\mu\nu\lambda\rho}\;\;
\sum_{i,j,k}\;\;\;
\sum_{p1,p2,p3,p4}\;\;\;
\left[{}^{ij}G_{\mu\nu}{{}^{p_1 p_3}_{p_2 p_2}}\right]\;
\left[{}^{ik}G_{\lambda\rho}{{}^{p_3 p_1}_{p_4 p_4}}\right]\, , 
\label{cla}
\eeq
\begin{eqnarray}
& & \W_3^B =
{1\over 2}\;\; B_{\mu\nu\lambda\rho}\;\;
B_{\alpha\beta\gamma\delta}\, 
\sum_{i,j,k}\;\;\;\; \sum_{p_1 , p_2 , p_3 , p_4}\;\;\;\;
\sum_{q_1 , q_2 , q_3 , q_4}\;\;\;\;
\nonumber\\
& &
\quad\qquad\;\,\left(
\left({}^{i}d_\mu\,\delta^{p_1 p_2} + {}^{i}p_\mu^{p_1 p_2}\right)\,
\left({}^{j}d_\alpha\,\delta^{q_1 q_2} + {}^{j}p_\alpha^{q_1 q_2}\right)\,
\left[{}^{ij}G_{\nu\beta}{{}^{p_2 q_3}_{q_2 p_3}}\right]\,
\left[{}^{jk}G_{\lambda\delta}{{}^{p_3 q_1}_{q_4 p_4}}\right]\,
\left[{}^{ki}G_{\rho\gamma}{{}^{p_4 q_4}_{q_3 p_1}}\right]
\right.
\nonumber\\
& &
\quad\qquad +\, \left({}^{i}d_\mu\,\delta^{p_1 p_2} + {}^{i}p_\mu^{p_1 p_2}\right)\,
\left({}^{k}d_\alpha\,\delta^{q_1 q_2} + {}^{k}p_\alpha^{q_1 q_2}\right)\,
\left[{}^{ij}G_{\nu\gamma}{{}^{p_2 q_4}_{q_3 p_3}}\right]\,
\left[{}^{jk}G_{\lambda\beta}{{}^{p_3 q_3}_{q_2 p_4}}\right]\, 
\left[{}^{ki}G_{\rho\delta}{{}^{p_4 q_1}_{q_4 p_1}}\right]
\nonumber\\
& &
\left. \quad\qquad +\, 
\left({}^{i}d_\mu\,\delta^{p_1 p_2} + {}^{i}p_\mu^{p_1 p_2}\right)\,
\left({}^{i}d_\alpha\,\delta^{q_1 q_2} + {}^{i}p_\alpha^{q_1 q_2}\right)\,
\left[{}^{ij}G_{\nu\delta}{{}^{p_2 q_1}_{q_4 p_3}}\right]\,
\left[{}^{jk}G_{\lambda\gamma}{{}^{p_3 q_4}_{q_3 p_4}}\right]\,
\left[{}^{ki}G_{\rho\beta}{{}^{p_4 q_3}_{q_2 p_1}}\right] 
\right)\, .
\nonumber\\
& &
\label{clb}
\end{eqnarray}
Our notation above for $\left[{}^{ij}G_{\mu\nu}{{}^{p_1 p_2}_{p_3 p_4}}\right]$
stands for the symmetrised bosonic Green's function
\beq
\left[{}^{ij}G_{\mu\nu}{{}^{p_1 p_2}_{p_3 p_4}}\right] =
{1\over 2}\, \left({}^{ij}G_{\mu\nu}{{}^{p_1 p_2}_{p_3 p_4}} + 
{}^{ji}G_{\nu\mu}{{}^{p_3 p_4}_{p_1 p_2}}\right)\, ,
\label{clc}
\eeq
and the coefficient $B_{\mu\nu\lambda\rho}$ is given
in (\ref{bi}). The sum over the indices 
$(p_1 , p_2 , p_3 , p_4)$ in (\ref{cla}) and (\ref{clb}) 
must respect of course the left-right multiplication property 
of $\left[{}^{ij}G_{\mu\nu}{{}^{p_1 p_2}_{p_3 p_4}}\right]$
and the order of the block indices $(i,j,k)$. The same field 
theory techniques lead to the evaluation of the cubic 
fermionic and ghost interactions which are respectively
found to be  
\begin{eqnarray}
\W_3^F & = & -\;{1\over 2}\; 
\sum_{i,j,k}\;\;\;
\sum_{p_2 , p_2 , p_3}\;\;
\sum_{q_1 , q_2 , q_3}\;\;
\sum_{a , b , c , d}\;\;
\left(
\left[{}^{ij}G_{\mu\nu}{{}^{p_3 q_1}_{q_3 p_1}}\right]\; 
\Gamma^{\mu}_{ab}\; 
\left({}^{ki}H_{bc}{{}^{p_2 q_2}_{q_1 p_3}}\right)\;
\Gamma^{\nu}_{cd}\;
\left({}^{kj}H_{da}{{}^{q_2 p_2}_{p_1 q_3}}\right)
\right.
\nonumber\\
& & 
\left.\qquad\qquad\qquad\qquad\quad\quad +\,
\left[{}^{ij}G_{\mu\nu}{{}^{p_3 q_1}_{q_3 p_1}}\right]\;
\Gamma^{\nu}_{ab}\; 
\left({}^{ik}H_{bc}{{}^{q_1 p_3}_{p_2 q_2}}\right)\;
\Gamma^{\mu}_{cd}\;
\left({}^{jk}H_{da}{{}^{p_1 q_3}_{q_2 p_2}}\right)
\right)\, ,
\nonumber\\
& & 
\label{cld}
\end{eqnarray}
\begin{eqnarray}
& & \W_3^{ghost} =
-\;\; 
\sum_{i,j,k}\;\;\;\; \sum_{p_1 , p_2 , p_3 , p_4}\;\;\;\;
\sum_{q_1 , q_2 , q_3 , q_4}\;\;\;\;
\nonumber\\
& &
\quad\qquad\;\,\left(
\left({}^{i}d_\mu\,\delta^{p_1 p_2} + {}^{i}p_\mu^{p_1 p_2}\right)\,
\left({}^{i}d_\nu\,\delta^{q_1 q_2} + {}^{i}p_\nu^{q_1 q_2}\right)\,
\left[{}^{kj}G_{\mu\nu}{{}^{p_3 q_4}_{q_3 p_4}}\right]\,
\left({}^{ji}E{{}^{p_4 q_3}_{q_2 p_1}}\right)\,
\left({}^{ki}E{{}^{q_4 p_3}_{p_2 q_1}}\right)
\right.
\nonumber\\
& &
\quad\qquad +\, 
\left({}^{i}d_\mu\,\delta^{p_1 p_2} + {}^{i}p_\mu^{p_1 p_2}\right)\,
\left({}^{i}d_\nu\,\delta^{q_1 q_2} + {}^{i}p_\nu^{q_1 q_2}\right)\,
\left[{}^{kj}G_{\mu\nu}{{}^{p_3 q_4}_{q_3 p_4}}\right]\,
\left({}^{ij}E{{}^{q_2 p_1}_{p_4 q_3}}\right)\,
\left({}^{ik}E{{}^{p_2 q_1}_{q_4 p_3}}\right)
\nonumber\\
&&\quad\qquad +\, 
\left({}^{k}d_\mu\,\delta^{p_1 p_2} + {}^{k}p_\mu^{p_1 p_2}\right)\,
\left({}^{i}d_\nu\,\delta^{q_1 q_2} + {}^{i}p_\nu^{q_1 q_2}\right)\,
\left[{}^{ik}G_{\mu\nu}{{}^{p_4 q_1}_{q_4 p_1}}\right]\,
\left({}^{ji}E{{}^{p_3 q_3}_{q_2 p_4}}\right)\,
\left({}^{jk}E{{}^{q_3 p_3}_{p_2 q_4}}\right)
\nonumber\\
&&\quad\qquad +\, 
\left({}^{k}d_\mu\,\delta^{p_2 p_3} + {}^{k}p_\mu^{p_2 p_3}\right)\,
\left({}^{j}d_\nu\,\delta^{q_2 q_3} + {}^{j}p_\nu^{q_2 q_3}\right)\,
\left[{}^{kj}G_{\mu\nu}{{}^{p_3 q_4}_{q_3 p_4}}\right]\,
\left({}^{ji}E{{}^{p_4 q_2}_{q_1 p_1}}\right)\,
\left({}^{ki}E{{}^{q_4 p_2}_{p_1 q_1}}\right)
\nonumber\\
&&\quad\quad -\,2\, 
\left({}^{i}d_\mu\,\delta^{p_1 p_2} + {}^{i}p_\mu^{p_1 p_2}\right)\,
\left({}^{j}d_\nu\,\delta^{q_2 q_3} + {}^{j}p_\nu^{q_2 q_3}\right)\,
\left[{}^{kj}G_{\mu\nu}{{}^{p_3 q_4}_{q_3 p_4}}\right]\,
\left({}^{ji}E{{}^{p_4 q_2}_{q_1 p_1}}\right)\,
\left({}^{ki}E{{}^{q_4 p_3}_{p_2 q_1}}\right)
\nonumber\\
&&\quad\quad \left. -\,2\,
\left({}^{k}d_\mu\,\delta^{p_1 p_2} + {}^{k}p_\mu^{p_1 p_2}\right)\,
\left({}^{j}d_\nu\,\delta^{q_2 q_3} + {}^{j}p_\nu^{q_2 q_3}\right)\,
\left[{}^{ik}G_{\mu\nu}{{}^{p_4 q_1}_{q_4 p_1}}\right]\,
\left({}^{ji}E{{}^{p_3 q_2}_{q_1 p_4}}\right)\,
\left({}^{jk}E{{}^{q_3 p_3}_{p_2 q_4}}\right)
\right)\, .
\nonumber\\
&&\label{cle}
\end{eqnarray}
As for the bosonic case, we need to keep track also here of 
the order of the block indices $(i,j,k)$ as well as the 
left-right multiplication of the propagators while summing 
over the $(p_1 , p_2 , p_3 , p_4)$ and $(q_1 , q_2 , q_3 , q_4)$ 
indices. The sum over the $(a,b,c,d)$ indices in (\ref{cld}) will 
turn into a trace over the product of the Dirac gamma matrices
where we can use the identities in (\ref{brx}).
\smallskip

\subsection{Comparison to the One-Loop Result}

To compare the contribution of the two-loop effective action 
to the one-loop result in (\ref{bw}) we proceed to replacing 
in the expressions of $\W_4^B$, $\W_3^B$, $\W_3^F$ and 
$\W_3^{ghost}$ above by the long-range expansion of the 
propagators as given by $(\ref{cg})$, $(\ref{ch})$ and 
$(\ref{ci})$. Using the rules of $(\ref{bs})$, $(\ref{bss})$,
$(\ref{bsss})$ and $(\ref{bt})$ and performing the sums over
different indices involved in the expressions of $\W_4^B$, 
$\W_3^B$, $\W_3^F$ and $\W_3^{ghost}$ above we can classify 
our results from the different contributions in the increasing 
power of $1/({}^id - {}^jd)^2$ as follows:
\begin{eqnarray}
& & \W_4^B = \sum_{i,j,k}\;\;
\left[
-\, {{45\, n_i\, n_j\, n_k}\over{\left({}^{ij}d\right)^2\,  
\left({}^{ik}d\right)^2}}
- \, {{36\, n_j\, n_k}\over{\left({}^{ij}d\right)^2\,  
\left({}^{ik}d\right)^6}}\, tr\left({}^{i}f_{\mu\nu}\,
{}^{i}f_{\mu\nu}\right)\right. 
\nonumber\\
& &
\quad\left.
- \, {{36\, n_j\, n_i}\over{\left({}^{ij}d\right)^2\,  
\left({}^{ik}d\right)^6}}\, tr\left({}^{k}f_{\mu\nu}\,
{}^{k}f_{\mu\nu}\right) 
-\, {{6\, n_j\, n_k}\over{\left({}^{ij}d\right)^4\,  
\left({}^{ik}d\right)^4}}\, tr\left({}^{i}f_{\mu\nu}\,
{}^{i}f_{\mu\nu}\right)
+\, \O\left({1\over d^9}\right)\right]\, .
\label{cma} 
\end{eqnarray}
\begin{eqnarray}
& & \W_3^B = \sum_{i,j,k}\;\;
\left[
{{27\, n_i\, n_j\, n_k}\over{2\,\left({}^{ij}d\right)^2\,  
\left({}^{ik}d\right)^2}}
\,+ \, {{2\, n_j\, n_k}\over{\left({}^{ij}d\right)^4\,  
\left({}^{ik}d\right)^4}}\, tr\left({}^{i}f_{\mu\nu}\,
{}^{i}f_{\mu\nu}\right)
+\, \O\left({1\over d^9}\right)\right]\, ,
\label{cmb} 
\end{eqnarray}
\begin{eqnarray}
& & \W_3^F = \sum_{i,j,k}\;\;
\left[
{128\,{n_i\,n_j\, n_k}\over{\left({}^{ij}d\right)^2\,  
\left({}^{ik}d\right)^2}}
+ \, {{768\, n_j\, n_k}\over{\left({}^{ij}d\right)^2\,  
\left({}^{ik}d\right)^4\,\left({}^{jk}d\right)^4 }}\,
tr\left({}^{i}f_{\mu\nu}\,
{}^{i}f_{\mu\nu}\right)\, 
+\, \O\left({1\over d^9}\right)\right]\, ,
\label{cmc} 
\end{eqnarray}
and finally the expression of $\W_3^{ghost}$ is also 
calculated up to the $1/d^8$ order and is exactly found to be
given by: $\W_3^{ghost} = - \W_4^B - \W_3^B - \W_3^F +\O(/1d^9)$.
This establishes our claim in this paper.
\smallskip

\section{Conclusions}

In this paper we derived the effective action for the IKKT
matrix model up to two loops for the scattering of an
arbitary number D-brane objects of the type IIB string theory.
The one-loop computation in [22] revealed the $F^4 / r^8$
behavior of the effective action. Our calulation at two-loops
showed that no renormalization of the $1/r^8$-term in the
effective potential ocur. 
\smallskip

These results are in agreement with the arguments made
in [22] that $1/r^8$ begavior in the efefctive interactions
ensures the cluster property among the D-objects which is
important to the $N=2$ supersymmetry of the IKKT action
and also to the dynamical generation of the spacetime 
coordinates. Furthermore, the exact agreement between the
IKKT matrix model and the supergravity results (lon-range
interactions) is an importanat consistency check critreion
for the IKKT matrix model to describe type IIB
superstrings.
\smallskip
 
\section{Appendix}

As advertised in Section~(4.1), we shall below present
some examples where we show how we take into account 
the fact that the propagators ${}^{ij}G_{\mu\nu}{{}^{qr}_{sp}}$, 
${}^{ij}H_{ab}{{}^{qr}_{sp}}$ and ${}^{ij}E{{}^{qr}_{sp}}$ act 
from the left and from the right while summing over the different 
indices appearing in $\W_4^B$, $\W_3^B$, $\W_3^F$ and
$\W_3^{ghots}$ given in (\ref{cla}), (\ref{clb}), (\ref{cld})
and (\ref{cle}), respectively. Since in this paper, we are only 
interested in the long-range contributions of the two-loop 
vacuum graphs we can proceed by replacing $1/P^2$ by $1/d^2$ in all 
the propagators. After doing this, the $1/d^2$ will act simply
as an overall real number factor but the field strength $F_{\mu\nu}$ 
keeps its action from the left and from the right while summing over 
the indices. To illustrate this more, we start by putting indices on 
the bosonic Green's function which are given in (\ref{cg}) as they 
appear in the term $\W_4^B$ of the two-loop effective action.
For example, we have for ${}^{ij}G_{\mu\nu}{{}^{qr}_{sp}}$
\begin{eqnarray}
{}^{ij}G_{\mu\nu}{{}^{qr}_{sp}}
& = & {1\over{\left({}^{ij}d\right)^2}}\,
\delta_{\mu\nu}\,
\delta^{qr}\, \delta^{sp} + 2\, i\,
{1\over{\left({}^{ij}d\right)^4}}\,
\left[{}^{i}f_{\mu\nu}{}^{qr}\, \delta^{sp}\, - \,
{}^{j}f_{\mu\nu}{}^{sp}\, \delta^{qr}\right]
\nonumber\\
& & -\; 4\, {1\over{\left({}^{ij}d\right)^6}}\,
\left[\left({}^{i}f_{\mu\alpha}\,
{}^{i}f_{\alpha\nu}\right)^{qr}\,
\delta^{sp}\,
+\, \left({}^{j}f_{\mu\alpha}\,
{}^{j}f_{\alpha\nu}\right)^{sp}\, \delta^{qr}\right.
\nonumber\\
&&\qquad\left.
-\, {}^{i}f_{\mu\alpha}{}^{qr}\,
{}^{j}f_{\alpha\nu}{}^{sp}\;
 -\;
{}^{j}f_{\mu\alpha}{}^{sp}\,
{}^{i}f_{\alpha\nu}{}^{qr}\right]\,
\; +\; \O\left(F^3\, ,\, 1/d^8\right)\, .
\label{za}
\end{eqnarray}
If we set $q=r$ and sum over $q$ using the fact that 
$tr\left({}^{i}f_{\mu\nu}\right)=0$, the above expression 
simplifies to
\begin{eqnarray}
\sum_q {}^{ij}G_{\mu\nu}{{}^{qq}_{sp}}
& = & {1\over{\left({}^{ij}d\right)^2}}\,
\delta_{\mu\nu}\,
n_i\, \delta^{sp}\,-\, 2\, i\,
{1\over{\left({}^{ij}d\right)^4}}\,
n_i\;\, {}^{j}f_{\mu\nu}{}^{sp}
\nonumber\\
&& -\; 4\, {1\over{\left({}^{ij}d\right)^6}}\,
\left[tr\left({}^{i}f_{\mu\alpha}\,
{}^{i}f_{\alpha\nu}\right)\,
\delta^{sp}\,
+\, n_i\, \left({}^{j}f_{\mu\alpha}\,
{}^{j}f_{\alpha\nu}\right)^{sp}\right]\; .
\label{zb}
\end{eqnarray}
Similarly, if we set $s=p$ and sum over $p$
using $tr\left({}^{j}f_{\mu\nu}\right)=0$, we have
\begin{eqnarray}
\sum_p {}^{ij}G_{\mu\nu}{{}^{qr}_{pp}}
& = & {1\over{\left({}^{ij}d\right)^2}}\,
\delta_{\mu\nu}\,
n_j\, \delta^{qr}\,+\, 2\, i\,
{1\over{\left({}^{ij}d\right)^4}}\,
n_j\;\, {}^{i}f_{\mu\nu}{}^{qr}
\nonumber\\
&& -\; 4\, {1\over{\left({}^{ij}d\right)^6}}\,
\left[tr\left({}^{j}f_{\mu\alpha}\,
{}^{j}f_{\alpha\nu}\right)\,
\delta^{qr}\,
+\, n_j\, \left({}^{i}f_{\mu\alpha}\,
{}^{i}f_{\alpha\nu}\right)^{qr}\right]\; .
\label{zc}
\end{eqnarray}
Further simplifications occur if we set $q=r$ and $s=p$
in(\ref{za}) and summing over $q$ and $p$ afterwards. 
The end result is then
\begin{eqnarray}
\sum_q \sum_p {}^{ij}G_{\mu\nu}{{}^{qq}_{pp}}
& = & {{n_i\, n_j\,\delta_{\mu\nu}}\over{\left({}^{ij}d\right)^2}}\,
\, -\; {4\over{\left({}^{ij}d\right)^6}}\,
\left[n_j\, tr\left({}^{i}f_{\mu\alpha}\,
{}^{i}f_{\alpha\nu}\right)\,
+\, n_i\, tr\left({}^{j}f_{\mu\alpha}\,
{}^{j}f_{\alpha\nu}\right)\right]\; .
\label{zd}  
\end{eqnarray}
\smallskip

From the above remarks, the expression of
$\W_4^B$ in (\ref{cma}) follows readily since 
it is a simple product of two Green's functions 
of the type (\ref{zb}) and (\ref{zc}) with the 
approriate sum over the block indices as required 
by the trace operation. The evaluation of $\W_3^B$
uses the same manipulations as $\W_4^B$ except that
it is more tedious since it involves the contraction 
of three Green's fucntions. The other expressions of
$\W_3^F$ and $\W_3^{ghots}$ are also easy to derive
once we know how to put and contract the indices in the
fermionic and ghost propagators. As for the bosonic
propagator above, we display here only the main
formulas that are needed. For the ghost propagator they are
\begin{eqnarray}
{}^{ij}E{{}^{qr}_{sp}} & = &
{{\delta^{qr}\,\delta^{sp}}\over{\left({}^{ij}d\right)^2}}\; , 
\label{ua}\\
\sum_q {}^{ij}E{{}^{qq}_{sp}} & = &
{{n_i\,\delta^{sp}}\over{\left({}^{ij}d\right)^2}}\; ,
\label{ub}\\
\sum_p {}^{ij}E{{}^{qr}_{pp}} & = &
{{n_j\,\delta^{qr}}\over{\left({}^{ij}d\right)^2}}\; ,
\label{uc}\\
\sum_q \sum_p {}^{ij}E{{}^{qr}_{pp}} & = &
{{n_i\, n_j}\over{\left({}^{ij}d\right)^2}}\; ,
\label{ud}
\end{eqnarray}
and for the fermionic propagator they are
\begin{eqnarray}
{}^{ij}H_{ab}{{}^{qr}_{sp}} & = & 
-\, {}^{ij}d\!\!\!/_{ab}\, 
{{\delta^{qr}\,\delta^{sp}}\over{\left({}^{ij}d\right)^2}}\, 
+\, {i\over 2}\, {1\over{\left({}^{ij}d\right)^4}}\,
\left({}^{ij}d\!\!\!/\, \Gamma^{\mu\nu}\right)_{ab}\, 
\left[{}^{i}f_{\mu\nu}{}^{qr}\, \delta^{sp}\, - \,
{}^{j}f_{\mu\nu}{}^{sp}\, \delta^{qr}\right]\, 
\nonumber\\
&& +\; {1\over 4}\,  {1\over{\left({}^{ij}d\right)^6}}\, 
\left({}^{ij}d\!\!\!/\,
\Gamma^{\mu\nu}\,\Gamma^{\alpha\beta}\right)_{ab}\, 
\left[\left({}^{i}f_{\mu\nu}\,
{}^{i}f_{\alpha\beta}\right)^{qr}\,
\delta^{sp}\,
+\, \left({}^{j}f_{\mu\nu}\,
{}^{j}f_{\alpha\beta}\right)^{sp}\, \delta^{qr}\right.
\nonumber\\
&&\qquad\left.
-\, {}^{i}f_{\mu\nu}{}^{qr}\,
{}^{j}f_{\alpha\beta}{}^{sp}\; -\;
{}^{j}f_{\mu\nu}{}^{sp}\,
{}^{i}f_{\alpha\beta}{}^{qr}\right]
+ \O\left(F^3 , 1/d^7\right)\; ,
\label{va}
\end{eqnarray}
\begin{eqnarray}
\sum_q {}^{ij}H_{ab}{{}^{qq}_{sp}} & = &
-\, {}^{ij}d\!\!\!/_{ab}\, 
{{n_i\,\delta^{sp}}\over{\left({}^{ij}d\right)^2}}\,
-\, {i\over 2}\, {1\over{\left({}^{ij}d\right)^4}}\,
\left({}^{ij}d\!\!\!/\, \Gamma^{\mu\nu}\right)_{ab}\;
n_i\, {}^{j}f_{\mu\nu}{}^{sp}\,
\nonumber\\
&& +\; {1\over 4}\,  {1\over{\left({}^{ij}d\right)^6}}\,
\left({}^{ij}d\!\!\!/\, 
\Gamma^{\mu\nu}\,\Gamma^{\alpha\beta}\right)_{ab}\,
\left[tr\left({}^{i}f_{\mu\nu}\,
{}^{i}f_{\alpha\beta}\right)\,
\delta^{sp}\,
+\, n_i\, \left({}^{j}f_{\mu\nu}\,
{}^{j}f_{\alpha\beta}\right)^{sp}\right],
\label{vb}  
\end{eqnarray}
\begin{eqnarray}
\sum_p {}^{ij}H_{ab}{{}^{qr}_{pp}} & = &
-\, {}^{ij}d\!\!\!/_{ab}\,
{{n_j\,\delta^{qr}}\over{\left({}^{ij}d\right)^2}}\,
+\, {i\over 2}\, {1\over{\left({}^{ij}d\right)^4}}\,
\left({}^{ij}d\!\!\!/\, \Gamma^{\mu\nu}\right)_{ab}\;
n_j\, {}^{i}f_{\mu\nu}{}^{qr}\,
\nonumber\\
&& +\; {1\over 4}\,  {1\over{\left({}^{ij}d\right)^6}}\,
\left({}^{ij}d\!\!\!/\,
\Gamma^{\mu\nu}\,\Gamma^{\alpha\beta}\right)_{ab}\,
\left[tr\left({}^{j}f_{\mu\nu}\,
{}^{j}f_{\alpha\beta}\right)\,
\delta^{qr}\,
+\, n_j\, \left({}^{i}f_{\mu\nu}\,
{}^{i}f_{\alpha\beta}\right)^{qr}\right],
\label{vc}
\end{eqnarray}
\begin{eqnarray}
\sum_q \sum_p {}^{ij}H_{ab}{{}^{qq}_{pp}} & = &
-\, 
{{{}^{ij}d\!\!\!/_{ab}\, n_i\, n_j}\over{\left({}^{ij}d\right)^2}}
\,+\; 
{{\left({}^{ij}d\!\!\!/\,
\Gamma^{\mu\nu}\,\Gamma^{\alpha\beta}\right)_{ab}\, 
}\over{4\, \left({}^{ij}d\right)^6}}\,
\left[n_i\, tr\left({}^{j}f_{\mu\nu}\,
{}^{j}f_{\alpha\beta}\right)\,
+\, n_j\, tr\left({}^{i}f_{\mu\nu}\,
{}^{i}f_{\alpha\beta}\right)\right].
\nonumber\\
&&
\label{vd}
\end{eqnarray}
The calculation of $\W_3^F$ involves an extra step
which consists of tracing over the Dirac matrices
using the identities in (\ref{brx}).

\pagebreak

\input epsf.tex
\begin{figure}
\centerline{\epsfxsize 17cm \epsfbox {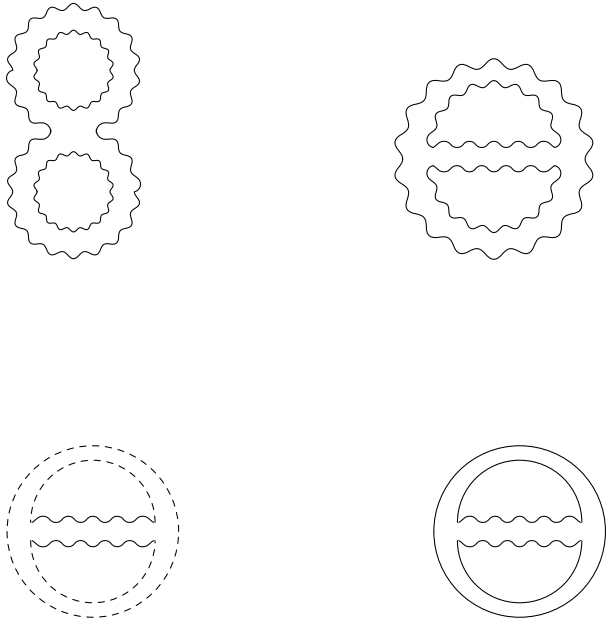}}
\vskip -6truecm
\caption{The standard four two-loop vacuum graphs. 
Wavy lines are gauge field matrix propagators. 
Solid lines are the propagators for the fermionic 
quantum fluctuation matrices and the dashed lines 
are the matrix ghost propagators.}
\end{figure}

\pagebreak

\section*{Acknowledgments}

This research was supported by the Institut des Hautes Etudes
Scientifiques (IHES) at Paris and the Abdus Salam Centre of 
Theoretical Physics (ICTP) at Trieste. The author is grateful 
to thank both IHES and ICTP for their hospitality and support 
while this work was being done. It is a pleasure to 
acknowledge a collaboration with A. Schwarz and M. R. Douglas 
at the earlier stage of this work. We would like also to 
thank C. Bachas, I. Antoniadis, D. Polyakov and A. Recknagel
for useful conversations and N. Ishibashi and M. R. Douglas 
for reading my paper and fruitful correspondence. Finally, 
I would like to thank P. A. Tirao at ICTP for assistance in 
the making of Figure~1 of this paper. 

\pagebreak

\end{document}